\documentclass[aps,showpacs,showkeys,preprintnumbers,amsmath,amssymb,nofootinbib]{revtex4}

\usepackage{amsmath}
\usepackage{amsfonts}
\usepackage{amssymb}
\usepackage{subfigure}

\usepackage{graphicx}
\usepackage{color}

\def \be  {\begin{equation}}
\def \ee  {\end{equation}}
\def \ee  {\end{equation}}
\def \bea {\begin{eqnarray}}
\def \eea {\end{eqnarray}}

\begin{document}

\preprint{ECTP-2016-01}
\preprint{WLCAPP-2016-01}
\hspace*{3mm}

\title{On dynamical net-charge fluctuations within a hadron resonance gas approach}

\author{Abdel Nasser Tawfik\footnote{The authors declare that there is no conflict of interest regarding the publication of this manuscript.}}
\email{a.tawfik@eng.mti.edu.eg}
\affiliation{Egyptian Center for Theoretical Physics (ECTP), Modern University for Technology and Information (MTI), 11571 Cairo, Egypt}
\affiliation{World Laboratory for Cosmology and Particle Physics (WLCAPP), Cairo, Egypt}

\author{L. I. Abou-Salem}
\affiliation{Physics Department, Faculty of Science, Benha University, 13518, Benha, Egypt}

\author{Asmaa G. Shalaby}
\affiliation{Physics Department, Faculty of Science, Benha University, 13518, Benha, Egypt}

\author{M. Hanafy}
\affiliation{Physics Department, Faculty of Science, Benha University, 13518, Benha, Egypt}

\date{\today}

\begin{abstract}

The dynamical net-charge fluctuations (${\nu}_{dyn}$) in different particle ratios $K/{\pi}$, $K/p$, and $p/{\pi}$ are calculated from the hadron resonance gas (HRG) model and compared with STAR central Au+Au collisions at $\sqrt{s_{NN}}=7.7-200~$GeV and NA49 central Pb+Pb collisions at $\sqrt{s_{NN}}=6.3-17.3~$GeV. The three charged-particle ratios ($K/{\pi}$, $K/p$, and $p/{\pi}$) are determined as total and average of opposite and average of same charges. We find an excellent agreement between the HRG calculations and the experimental measurements, especially from STAR beam energy scan (BES) program, while the strange particles in the NA49 experiment at lower Super Proton Synchrotron (SPS) energies are not reproduced by the HRG approach. We conclude that the utilized HRG version seems to take into consideration various types of correlations including strong interactions through the heavy resonances and their decays especially at BES energies. 

\end{abstract}


\maketitle


\section{Introduction}

The heavy-ion experiments are designed to study the nuclear matter under extreme conditions of high temperature or density or both of them \cite{r1}.  A new state of matter called quark-gluon plasma (QGP), which is conjectured to shape the cosmic background geometry in a few microseconds after the Big Bang, is assumed to be created in heavy-ion collisions \cite{r2} and predicted by the quantum chromodynamics (QCD) \cite{r3}. The quark-hadron phase transition can be characterized by many experimental signatures \cite{r4,r5,r7,r8,r9,r10,r11,r12} such as the net-charge fluctuations of various produced particles \cite{r10,r12,r13,r14}. 
  
The net-charge fluctuations can be measured \cite{r13,r14,r15,r16,r17,r18} and calculated \cite{r9,r10,r11} in terms of different quantities such as the variance of the charge $V(Q)$, the variance of the charge ratios $V(R)$, the fluctuation of charged-particle ratios $D$ and  the dynamical net-charge fluctuations $\nu_{dyn}$. $V(Q)$, $V(R)$ and $D$ are affected by the measurement conditions such as the detector acceptance, the global charge conservation, the charge asymmetry, the background contributions and the random efficiency loss \cite{r10,r19}. $\nu_{dyn}$ is the only quantity which can be measured independent on the detector acceptance, see for instance Ref. \cite{r10}. Furthermore, $\nu_{dyn}$ is suitable for the calculations with antisymmetric charge distribution and the conservation of global charge \cite{r19}.   
  
For the sake of completeness, we mention that in relativistic heavy-ion collisions, the event-by-event dynamical charge, the baryon number and the strangeness fluctuations have been intensively investigated \cite{r8}. It was proposed that the dynamical charge fluctuations of positively- to negatively-charged pions give a measure for the number of rho- and omega-resonances which are likely produced through the hadronization process (quark-hadron phase transition) \cite{r8}. Therefore, the dynamical charge fluctuations have been proposed as a signature for the QGP formation \cite{r9} because the charge fluctuations are directly proportional to the square of the electric charge. In case of stable QGP, the charge fluctuations are slightly smaller than that in the hadron phase, due to the conservation of the charge fluctuations normalized to the entropy. Again, this signature was proposed because the fluctuations taking place in QGP are likely to be reduced through the rapid hadronization process \cite{r9,r20}. Also, processes such as rescattering and resonance decay may weaken such fluctuations. The resonance decay which occurs after hadronization and stops at the stage of chemical freezeout should principally be distinguished from the resonance decay which shall be taken into consideration in the HRG approach. Furthermore, the dynamical charge fluctuations in QGP  era are remarkably reduced from the ones characterizing the early stages of the heavy-ion collision which can be described by quarks and gluons degrees of freedom \cite{r21}. This illustrates the importance of considering the propagation of fluctuations from the very early stages of the collisions to the state of chemical freezeout via hadronization and other processes before drawing any conclusions about these fluctuations  as a reflection of the QGP formation. The temporal evolution of this propagation can be evaluated by means of transport approaches, for instance. In the present work, we utilize a statistical approach, which characterizes very well the produced particles in their final state \cite{r22}. The study of net-charge dynamical fluctuations - among others - are experimental tools to probe the final state of QCD processes.

In principle, the fluctuations can be estimated from statistical approaches as variance, covariance or higher-order moment \cite{r22}. The dependence of  mean transverse momentum and balance fluctuations on momentum is an essential tool to measure the fluctuations \cite{r23}. Any possible difference between the calculations and the measurements can be attributed to certain novel dynamics. These correlation functions are emerged as identical pion Hanbury Brown-Twiss (HBT) correlations \cite{r24}.

So far, the charge fluctuations have been analysed in several experiments. The PHENIX experiment has found that the charge fluctuations are consistent with that in hadron resonance gas (HRG) if the latter are extrapolated to larger detector acceptance, while the earlier are measured with a small rapidity acceptance \cite{r17}. The STAR experiment confirmed the PHENIX findings \cite{r7}. At SPS energies, both CERES \cite{r17} and NA49 \cite{r25} have reported consistency with pure pion gas predictions. Recently, the STAR experiment \cite{r26} measured the dependence of the dynamical net-charge fluctuations of $K/{\pi}$, $K/p$, and $p/{\pi}$ on the beam energy scan (BES) energies, $\sqrt{s_{NN}}=7.7-200~$GeV \cite{r7}. The variance of the net-charge fluctuations has been estimated as a function of multiplicity, azimuthal coverage of the detector, and centrality \cite{r17} and are given as functions of the collision participants and the pseudo-rapidity, as well \cite{r13}. 
 
We present HRG calculations for the energy dependence of the dynamical net-charge fluctuations $\nu_{dyn}$ in different particle ratios; $K/{\pi}$, $p/{\pi}$, and $K$/$p$. We have considered three cases, total charge ratios, average of the same and average of the opposite signs. We focus on $\nu_{dyn}$ as it is considered as a clean experimental signature not suffering from problems such as global charge conservation, charge asymmetry, background contributions, random efficiency loss and volume fluctuations.  
	
The present paper is organized as follows. The formalism of the HRG model is introduced in section \ref{sec:hrg}. Section \ref{sec:ncf} elaborates the net-charge fluctuations in particle ratios. The results and discussion of the net-charge fluctuations in $K/{\pi}$, $K/p$, and $p/{\pi}$ are presented in section \ref{sec:reslts}. In section \ref{sec:conls}, the final conclusions are outlined.

\section{Formalism}
\label{sec:forms}

\subsection{Hadron resonance gas model}
\label{sec:hrg}

The thermodynamic properties can be directly derived from the partition function $Z(\beta,\mu,V)$. In a grand canonical ensemble \cite{r27}
 \begin{equation} \label{GrindEQ__1_}
Z(\beta,\mu,V)= \mbox{Tr} \left\{ \exp \left[\left({\mu}N-H\right)\beta \right] \right\},
\end{equation}
where $\beta=1/T$ is the inverse of the temperature ($T$), $H$ represents the Hamiltonian of the system, $N$ is the number of particles, $V$ is the volume of the interacting system and $\mu$ is the chemical potential. The Hamiltonian is included as it contains all relevant degrees of freedom of the confined and the strongly interacting medium and implicitly includes other types of interactions which result in the resonances formation \cite{r27,r28,r29,r30,r31}. The grand canonical ensemble has two important features; the kinetic energies and the summation over all degrees of freedom and energies of the hadron resonances. As Hagedorn proposed, the formation of resonances can only be achieved through strong interactions, i.e. {\it "resonances (fireballs) are composed of further resonances (fireballs), which in turn consist of  resonances (fireballs) and so on".} This simply means that the contributions of the various hadron resonances to the total partition function are the same as those of ideal collisionless constituents with an effective mass.  At a temperature compatible with the resonance half-width, the effective mass approaches the physical one \cite{r3}. This implies that at high temperatures even strong interactions are conjectured to be taken into consideration through including heavy resonances. 
  
Therefore, the partition function, Eq. \eqref{GrindEQ__1_}, of a hadron resonance gas can be summed over the number of the hadron resonances composing the HRG degrees of freedom \cite{r32}
 \begin{equation} \label{GrindEQ__2_}
\ln Z(\beta,\mu,V)=\sum_i{{\ln Z}_i(\beta,\mu,V)},
\end{equation}
where $\mu\in[\mu_{B}, \mu_{S}, \mu_{Q}]$ are the chemical potentials related to the baryon number, the strangeness and the electric charge, respectively. The chemical potentials ($\mu$) can be related to the nucleus-nucleus center-of-mass energy ($\sqrt{s_{NN}}$), phenomenologically  \cite{r22,Tawfik:2013bza,Tawfik:2014dha},
\bea
\mu &=& \frac{a}{1+b\, \sqrt{s_{NN}}}, \label{eq:musqrts}
\eea
where $a=1.245\pm 0.094~$GeV and $b=0.264\pm 0.028~$GeV$^{-1}$. 

It should be highlighted that the hadron resonances with masses up to $2\;$GeV include a suitable set of constituents needed for the partition function. The concrete number of constituents does not matter. Nevertheless, this can be determined from the latest update (2014) of the particle data group. We have included all measons and baryons having masses $\leq 2~$GeV. Such a mass cut is defined in order to avoid the singularity expected at the Hagedorn temperature~\cite{r27,r28,r29,r30}. The strong interactions are assumed to be taken into consideration. In light of this, the validity of HRG is limited to temperatures below the critical one, $T_c$. 

In high energy experiments, the produced particles and their correlations and fluctuations are believed to provide essential information about the nature, size and composition of the QCD matter from which they are stemming. The freezeout parameters have been determined at various center-of-mass energies $\sqrt{s_{NN}}$. These are  temperature $T$ and baryon chemical potential $\mu$. The chemical freezeout is a stage in which the inelastic collisions entirely cease and the relative particle yields and ratios get fixed. As given in Eq. (\ref{eq:musqrts}), $\mu$ is related to $\sqrt{s_{NN}}$ (also $T$ can be given in dependence on $\sqrt{s_{NN}}$). $T$ vs. $\mu$ has been phenomenologically described, the so-called freezeout conditions, for instance, $s/T^3=7$ \cite{r22}. Accordingly, temperatures and baryon chemical potentials can be determined for each beam energy, for instance Eq. (\ref{eq:musqrts}). Further interrelations can be taken from Ref. \cite{r22,Tawfik:2013bza,Tawfik:2014dha}.

The $i$-th hadron (resonance) partition function can be written as
\begin{equation} \label{GrindEQ__3_}
{\ln Z}_i\left(\beta,\mu,V\right)=\pm\frac{Vg_i}{2{\pi }^2}\int^{\infty }_0 k^2\ dk\, \ln  [1\pm {\lambda }_i\ {\rm exp}(-\beta {\varepsilon }_{i}(k))], 
\end{equation}
where $\pm$ stand for fermions and bosons respectively, $\varepsilon _{i}(k) =\sqrt{k^{2} +m_{i}^{2}}$ and the fugacity factor ${\lambda}_i$ \cite{r29} is given by
\begin{equation} \label{GrindEQ__4_}
\lambda _{i} (\beta,\mu )=\exp\left[\left(B_{i} \mu _{B} + S_{i} \mu _{S} +Q_{i} \mu _{Q}\right)\beta\right],
\end{equation}
where $B_i$, $S_i$ and $Q_i$  are baryon, strange and charge quantum number of $i$-th hadron resonance.

The number density can be obtained as \cite{r32}
\begin{equation} \label{GrindEQ__6_}
n(\beta,\mu) 
=\sum_{i}\pm\frac{g_{i}}{2\pi^{2}}\int _{0}^{\infty }\frac{k^{2}\ dk}{\exp\left[\left({\mu}_{i} - {\varepsilon }_{i}(k)\right)\beta\right] \pm 1}
\end{equation}
In section \ref{sec:ncf}, the dynamical net-charge fluctuations shall be derived in the HRG approach.

In determining $n(\beta,\mu)$ as an example on all other thermodynamic quantities, we take into account contributions from possible decay channels. Thus for $m$-th species, the final particle number density is calculated as 
\begin{eqnarray}
n_m^{total}(\beta,\mu) &=& n_m^{direct}(\beta,\mu) + \sum_{n\neq m} B_{n\rightarrow m}\,  n_n^{decaying}(\beta,\mu) \label{eq:decay}
\end{eqnarray}
where $B_{n\rightarrow m}$ being the decay branching ratio of $n$-th into $m$-th hadron resonance and $n_n^{decaying}$ is the number density of $n$-th resonance that decays in the particle species of interest.

\subsection {Dynamical net-charge fluctuations}
\label{sec:ncf}

Predicting the phase transition can be achieved through different signatures such as the fluctuations of the net-charges of the produced particles. As discussed in previous sections, the net-charge fluctuations as a signature for QGP formation is best accessible through ${\nu}_{dyn}$ \cite{r10,r19}. 

The standard deviation which describes the fluctuations in $p/{\pi}$ ratio, for instance, is given as \cite{r33,r34,r35}
 \begin{equation}  \label{GrindEQ__7_}
 \sigma^{2}_{p/\pi}=\ \frac{\langle{N_p}^2\rangle -{\langle N_p\rangle}^2}{{\langle N_p\rangle}^2}+\frac{\langle{N_\pi}^2\rangle-{\langle N_\pi\rangle}^2}{{\langle N_\pi\rangle}^2}-2\frac{\langle N_p N_{\pi }\rangle-\langle N_p\rangle\langle N_{\pi}\rangle}{\langle N_p\rangle\langle N_{\pi}\rangle},
 \end{equation}
where $N_p$ and $N_{\pi}$ represent the number of charged protons and pions, respectively. The covariance, which measures how much two {\it random} variables change with each other. The variance is a special case of covariance if the variables are identical. The quantity $\langle N_p N_{\pi }\rangle$ gives the statistical average of the integrals or multiplicities of the simultaneous  correlation of the production of both $p$ and $\pi$.

Equation \eqref{GrindEQ__7_} can be rewritten as
 \begin{equation} \label{GrindEQ__8_}
 \sigma^{2}_{p/\pi}=\ \frac{\langle N_p(N_p-1)\rangle}{{\langle N_p\rangle}^2}+\frac{\langle N_{\pi }(N_{\pi }-1)\rangle}{{\langle N_{\pi }\rangle}^2}-2\frac{\langle N_pN_{\pi }\rangle}{\langle N_p\rangle\langle N_{\pi }\rangle}+\frac{1}{\langle N_p\rangle}+\frac{1}{\langle N_{\pi }\rangle}.
 \end{equation} 
The last two terms in Eq. \eqref{GrindEQ__8_} represent {\it static} net-charge fluctuations in Piosson's limit (uncorrelated particle production, i.e. just statistical). Then, the charge dependence of the dynamical net-charge fluctuations for $p/\pi$ charge-ratio  can be given in terms of second-order factorial moments \cite{Adamovich:1997sw,Adamovich:2001hi,Tawfik:2001aw}, $\langle N_p(N_p-1)\rangle$ and/or $\langle N_{\pi}(N_{\pi}-1)\rangle$ and the covariance $\langle N_p N_{\pi }\rangle$,
\begin{equation} \label{GrindEQ__11_}
\left.{\nu}_{dyn}\right|_{p/\pi} =\ \frac{\langle N_p(N_p-1)\rangle}{{\langle N_p\rangle}^2}+\frac{\langle N_{\pi }(N_{\pi }-1)\rangle}{{\langle N_{\pi }\rangle}^2}-2\frac{\langle N_p N_{\pi }\rangle}{\langle N_p\rangle\langle N_{\pi }\rangle}.
\end{equation}
The dynamical fluctuations are related to single- and two-particle multiplicities. Straightforwardly, the expressions for $K/{\pi}$ and $K/p$ ratios can be derived as in Eq. \eqref{GrindEQ__11_}. For more information about the net-charge fluctuations, the readers are kindly advised to consult Refs. \cite{r36,r37,r38,r39}.

The net-charge dynamical fluctuations ($\nu$) may be affected by uncertainties arising from volume fluctuations and thus the study of particle (charge) ratio fluctuations is assumed to largely eliminate these uncertainties. $\nu$ might be influenced by the charge conservation and simultaneously by the limited charge available on the onset on the collision \cite{r10}. The study of particle (charge) ratio $\nu$ gives a measure for the relative correlation strength of particle pairs. Negative $\nu$ signifies a dominant contribution from correlated particle pairs.

In the statistical thermal model, the main source of correlations is the resonance decay, i.e., the second term in Eq. (\ref{eq:decay}) is the only source of different-species correlations. For instance, a $\Delta$ baryon decaying into a nucleon and a  pion certainly results in proton-pion correlation. Similarly, the difference between $\langle N_p\, N_K \rangle$ and $\langle N_p \rangle \langle N_K \rangle$ must come from a strange baryon decay.

\section{Results and discussion}
\label{sec:reslts}

\begin{figure}[htb]
\centering
\includegraphics[width=8.cm]{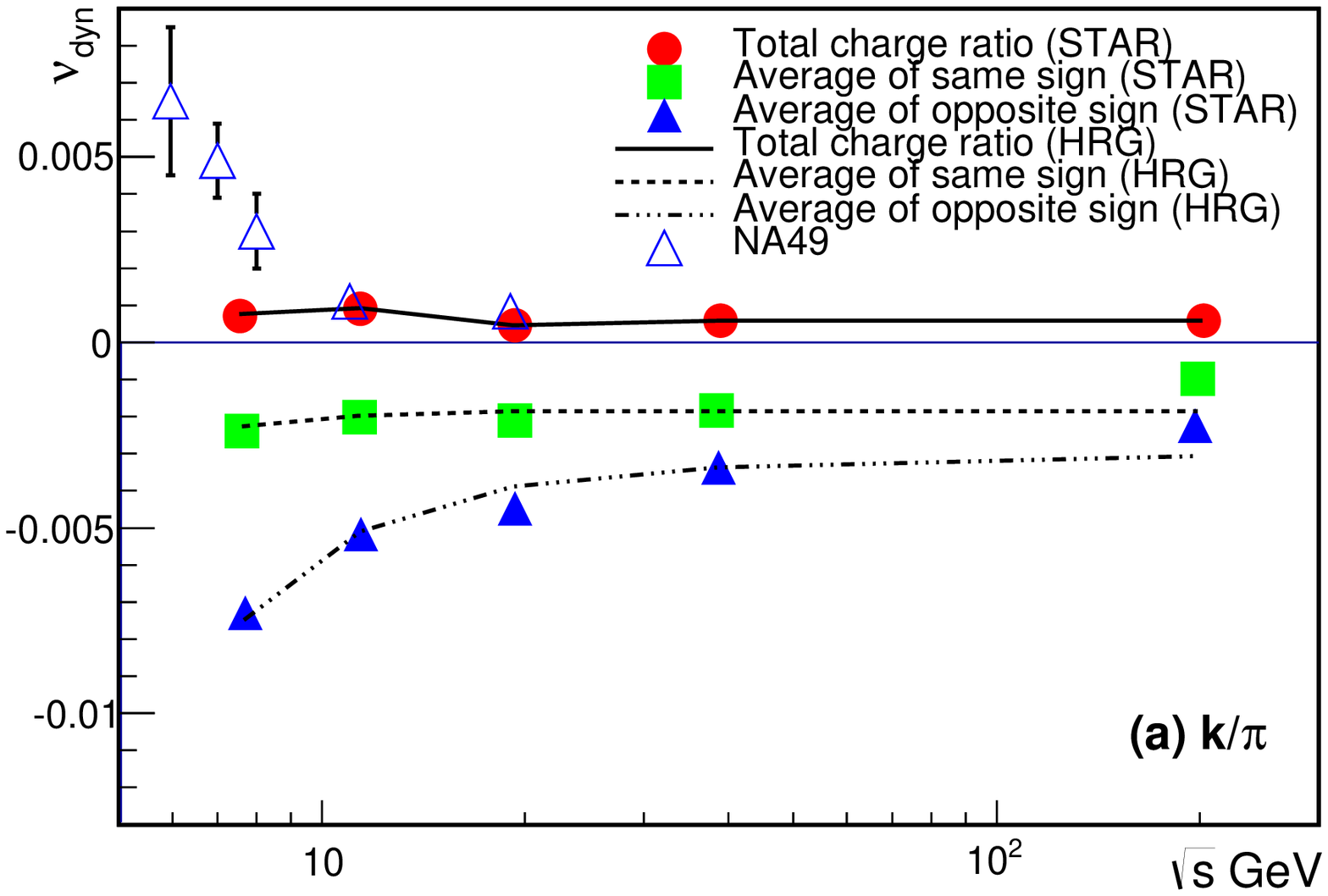} \\
\includegraphics[width=8.cm]{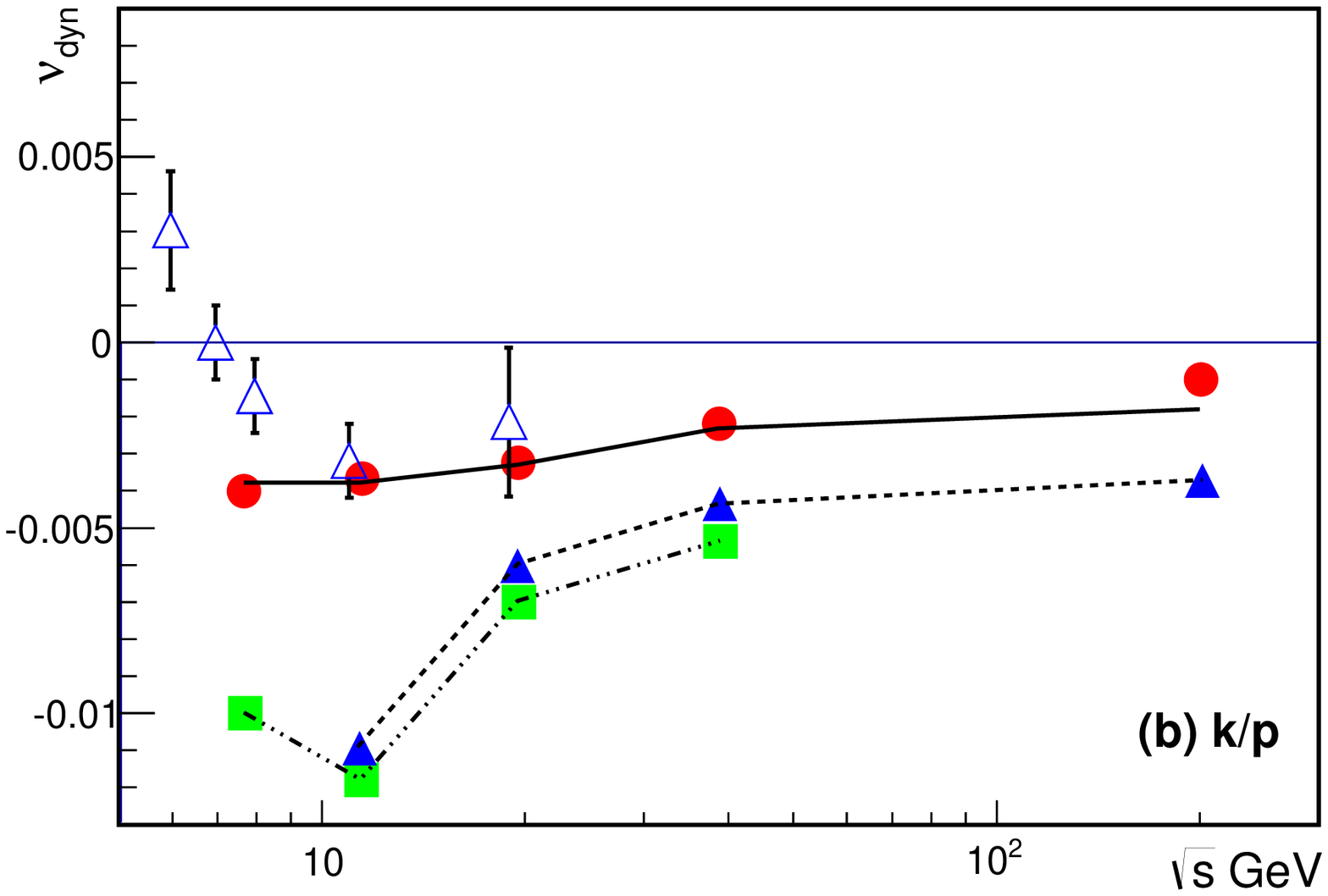} \\ 
\includegraphics[width=8.cm]{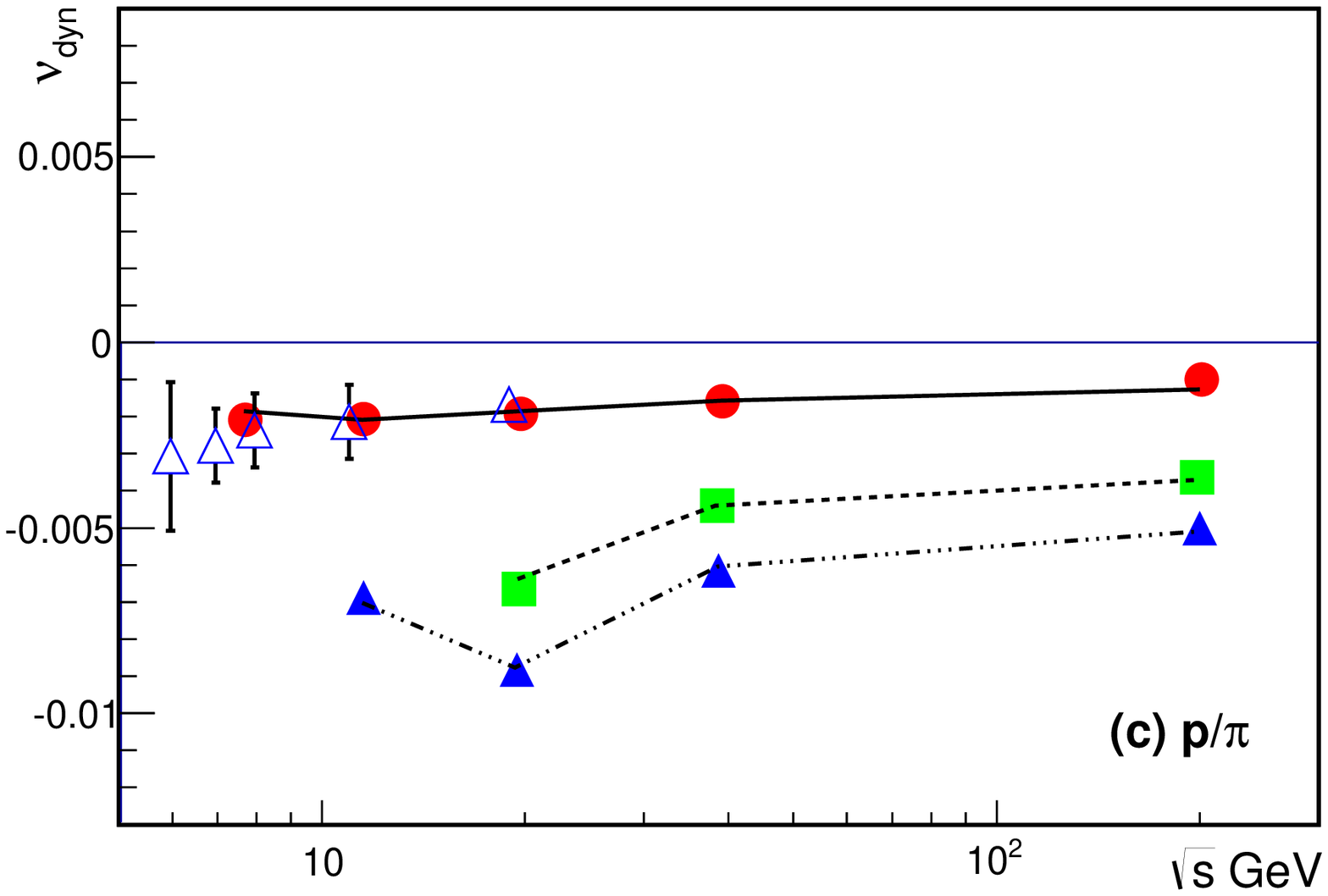}
\caption{The energy dependence of the dynamical net-charge fluctuations ${\nu}_{dyn}$ calculated from the HRG model is compared with STAR \cite{r26} (solid symbols) and NA49 \cite{r33} (open triangles) measurements for the charge-ratios $K/\pi$ (a), $K/p$ (b) and  $p/\pi$ (c). The fluctuations of the charge ratios are determined as total and average of same and average of opposite charge-signs.  }
\label{fig:new3}
\end{figure}

In Fig. \ref{fig:new3}, the dynamical net-charge fluctuations calculated from the HRG model are given as a function of beam energy for various charge-ratios $K/{\pi}$, $K/p$ and $p/{\pi}$. The HRG calculations are compared with STAR \cite{r26} (solid symbols) and NA49 \cite{r33} (open triangles) measurements. For each charge-ratio, three different cases are considered, namely the total dynamical fluctuations and the average of the same and of opposite charge-signs, which are characterized by solid, dashed, triple-dot-dashed curves, respectively. The charge-dependence of the dynamical net-charge fluctuations is conjectured to give detailed information about the correlations of various produced particles (charges) and on how they are correlated with each others.
  
The top panel of Fig. \ref{fig:new3} (a) shows ${\nu}_{dyn}$ for $K/{\pi}$ charge-ratio as a function of the beam energy. The HRG calculations are compared with STAR measurement in central $0-5\%$ Au+Au collisions at $\sqrt{s_{NN}}=7.7-200~$GeV \cite{r26} and NA49 for total charged particle ratios in central $0-3.5\%$ Pb+Pb collisions at $\sqrt{s_{NN}}=6.3-17.3~$GeV \cite{r33}. The three cases, namely dynamical fluctuations for total charges $(K^++K^-)/({\pi}^++{\pi}^-)$, average of same charges $K^+/{\pi }^+ + K^-/{\pi }^-$ and average of opposite signs $K^+/{\pi}^- + K^-/{\pi }^+$ are compared with each others. Only $(K^++K^-)/({\pi}^++{\pi}^-)$ charge-ratios have positive dynamical fluctuations, which remain constant with increasing the beam energy. Almost the same dependence is also observed in the average of same charges but with negative dynamical net-charge fluctuations. The average of opposite charges has a stronger energy dependence. It is obvious that the dynamical fluctuations of opposite-charged ratios have larger negative values, which decline with increasing the collision energy, than that of the same-charged ratios. The negative values for the dynamical net charge fluctuations can be understood from Eq. \eqref{GrindEQ__11_}. The last term (the correlation term) seems to have a larger value than the first two terms. Thus, the largest negative values obtained in opposite-charged particle ratios mean a remarkably larger cross-correlation of these two charges. 

The middle panel of Fig. \ref{fig:new3} (b) presents the same as in left-hand panel (a) but for $K/p$ charge-ratio. The dynamical fluctuations for total charges increase and then saturate with increasing $\sqrt{s_{NN}}$. The dynamical fluctuations of average of same-charged particles obviously increase with the energy. For oppositely charged particles-ratios the dynamical fluctuations first decrease and then increase with increasing the energy. Such a remarkable dependence needs a further confirmation at lower energies. Unfortunately, no experimental measurements are available so far. This might be possible with the Nuclotron-based Ion Collider fAcility (NICA) future facility at the Joint Institute for Nuclear Research (JINR) at Dubna-Russia and the Facility for Antiproton and Ion Research (FAIR) at the Gesellschaft fuer Schwerionenforsching (GSI) at Darmstadt-Germany. We observe that the values of the dynamical fluctuations of $K/p$ are always negative. This means that the produced kaons and protons are strongly correlated over all beam energies. 

The bottom panel of Fig. \ref{fig:new3} (c) depicts the same as in previous panels but for the $p/{\pi}$ charge-ratios. Their total-charge dynamical fluctuations slightly increase with the energy. Oppositely charged particles ratios first decrease and then increase with the energy. Again, further experimental measurements at lower energies shall play a crucial role in confirming this remarkable dependence. The dynamical fluctuations of same-signed $p/\pi$ increase with the energy. As in $K/p$ charged-particles ratios, here the dynamical fluctuations are always negative.
  
We have observed that the dynamical net-charge fluctuations possess a remarkable dependence on the beam energy, from which we follow the change in their values and thus might be able to signal the impact of QGP formation. The excellent agreement between the HRG calculations and the STAR measurements gives indication for the statistical origin of the particle production (charges). The excellent agreement reveals that the HRG fluctuations reproduce the ones for the produced particles which are conjectured to undergo a quark-hadron phase transition. The phase transition itself is not accessible by the HRG model due to absence of the underlying dynamics and missing of the degrees of freedom at least of the QGP phase. On the other hand, the produced particles in their final state (chemical freezeout) are likely affected by the QCD phase-transition(s). For sake of completeness, the version of the HRG model we have utilized in the present work \cite{r22} is conjectured to take into consideration various types of correlations including the strong interactions through the summation over heavy resonances \cite{r3}. While the HRG calculations are excellently consistent with the results from STAR experiment, they are not compatible with the NA49 measurements, especially for strange particles at lower SPS energies. Besides strangeness production at low energies, the Relativistic Heavy Ion Collider (RHIC) experiments are well suited for weak decay contributions and operate various specialized detectors which is not the case for the SPS experiments \cite{r40}.

\section{Conclusions}
\label{sec:conls}

In the present work, the dynamical net-charge fluctuations which are proposed as signatures of the QGP formation are calculated from ${\nu}_{dyn}$ by using the HRG approach,  in which the entire list of baryons and mesons with masses $\leq 2~$GeV from the latest particle data group are taken into account. This includes stable hadrons and unstable resonances. The latter can be decaying into the particles (charges) of interest and therefore their branching ratios are taken into consideration. For the sake of completeness, we emphasize that the constituents of the HRG models are assumed as point-like particles, i.e. no excluded-volume corrections were added. $\nu_{dyn}$ is considered as a suitable measurable quantity for the dynamical net-charge fluctuations. We have calculated the charge dependence of the dynamical net-charge fluctuations in $K/{\pi}$, $K/p$  and $p/{\pi}$ charge-ratios and compared the results with the STAR and NA49 corresponding measurements. An excellent agreement with the STAR BES results for the three charge-ratios at $\sqrt{s_{NN}}=7.7-200~$GeV is obtained. On the other hand, HRG agrees well with NA49 results at top SPS energies. But at lower SPS energies, HRG is not capable in reproducing NA49 dynamical net-charge fluctuations for strange particles, particularly. We found that only the $(K^++K^-)/({\pi}^++{\pi}^-)$ are positive. All other charge ratios and combinations of various charges result in negative net-charge fluctuations indicating that the cross term, last term in Eq. \eqref{GrindEQ__11_}, the correlated particle pair, is dominant against the first two terms (single {\it uncorrelated} particle). Thus, we conclude that the correlations between charge-ratios are larger than the dynamical fluctuations in the noncorrelated charges.  
  
To the authors' best knowledge this is the first time in which the HRG approach is implemented in calculating the dynamical net-charge fluctuations ($\nu_{dyn}$) for different charge-ratios as a function of energy and compared with STAR and NA49 experiments. We conclude that - at least - the HRG version utilized in the present work  (all particle data group's baryons and baryons with masses $\leq 2~$GeV and possible branching ratios (decay channels in the particles or charges of interest) but no excluded-volume corrections) seems to count for various types of correlations including the strong interactions obviously through the sum over heavy resonances, especially at BES energies. 
  



\end{document}